# The quest for identifiability in human functional connectomes

Enrico Amico[1,2] and Joaquín Goñi [1,2,3,*]
[1] School of Industrial Engineering, Purdue University, West-Lafayette, IN, USA
[2] Purdue Institute for Integrative Neuroscience, Purdue University, West-Lafayette, IN, USA
[3] Weldon School of Biomedical Engineering, Purdue University, West-Lafayette, IN, USA
[*] Correspondence: jgonicor@purdue.edu

## Abstract

The evaluation of the individual "fingerprint" of a human functional connectome (FC) is becoming a promising avenue for neuroscientific research, due to its enormous potential inherent to drawing single subject inferences from functional connectivity profiles. Here we show that the individual fingerprint of a human functional connectome can be maximized from a reconstruction procedure based on group-wise decomposition in a finite number of brain connectivity modes. We use data from the Human Connectome Project to demonstrate that the optimal reconstruction of the individual FCs through connectivity eigenmodes maximizes subject identifiability across resting-state and all seven tasks evaluated. The identifiability of the optimally reconstructed individual connectivity profiles increases both at the global and edgewise level, also when the reconstruction is imposed on additional functional data of the subjects. Furthermore, reconstructed FC data provide more robust associations with task-behavioral measurements. Finally, we extend this approach to also map the most task-sensitive functional connections. Results show that is possible to maximize individual fingerprinting in the functional connectivity domain regardless of the task, a crucial next step in the area of brain connectivity towards individualized connectomics.

## Introduction

The explosion of publicly available neuroimaging datasets in the last years have provided an ideal benchmark for mapping functional and structural connections in the human brain. At the same time, quantitative analysis of connectivity patterns based on network science have become more commonly used to study the brain as a network [1], giving rise to the area of research so called Brain Connectomics [2,3]. The analyses of functional and structural brain connectivity patterns have allowed researchers to make inferences on the different organization of brain networks in clinical and healthy populations, and to identify changes in these cohorts, usually by testing differences across groups [4,5]. Until recently, brain connectivity studies have generally overlooked the existing connectivity heterogeneity within each group, for several reasons. The group average procedure eases comparisons between different populations and has the benefit of providing more representative connectivity patterns. However, this comes at the cost of ignoring the potentially precious information provided by subject level, i.e. individual , connectomes.

Recent work on fMRI [6–9] and EEG fingerprinting [10] has paved the way to the new promising avenue of detecting individual differences through brain connectivity features. They showed that the individual functional connectivity (FC) profiles estimated from functional resonance magnetic imaging (fMRI) data can be seen as a "fingerprint" of the subject, which indeed may be used to identify a given individual in a set of functional connectivity profiles from a population.

The capacity of functional connectivity profiles to identify subjects goes along with the concept of reproducibility of test-retest experiments [6]. The rationale behind is that the higher the accuracy of the functional connectivity on each fMRI session and subject, the better will be the identifiability of individual subjects. Several aspects may have an impact on the quality, and hence on the identifiability of the data. This includes the characteristics of the fMRI sequence (such as its spatial and temporal resolution [11]), the processing of the fMRI data (including how to handle head motion and other artifacts [12,13] and the statistical approach used to obtain pairwise region-to-region functional connectivity from voxel-level time-series [4,14]. All the aspects listed above refer to efforts on increasing the reliability of functional connectivity on individual fMRI sessions [9,13,15]. Indeed, all of them could be applied by acquiring just one fMRI session of one subject. In the lack of gold-standards in brain connectivity, it is important to investigate the reliability of connectome fingerprinting [7,9,15] by procedures that gradually assess the data from common connectivity patterns present along the cohort to individual connectivity patterns.

The assessment of individual fingerprinting based on functional connectivity could be presented as a decomposition methodology. In this case, one could think of an approach that obtains common connectivity patterns highly present in the cohort, individual patterns present in certain subjects, and even in only certain individual sessions. Such a framework would allow us to decipher between what connectivity patterns are common in a cohort (human brain functional traits), what connectivity traits are unique of different individuals (inter-subject variability) and what are session-dependent (within-subject variability) or beyond, i.e. spurious patterns from the standpoint of individual fingerprinting. Interestingly, this conceptual framework would allow for individual reconstruction of functional connectomes based on a subset of the aforementioned traits.

Here we propose a group-level framework to assess and maximize the individual fingerprinting of functional connectomes based on a principal component analysis (PCA) decomposition and subsequent individual reconstruction. We show that the uniqueness of each individual connectivity profile can be reconstructed through an optimal finite linear combination of orthogonal principal components (or eigenmodes) in the connectivity domain, hence here denominated brain connectivity modes. These connectivity modes improve the identification of each individual's functional architecture both at the whole-brain and local sub-network level. We evaluate this methodology on 100 unrelated subjects of the Human Connectome Project (HCP), for test-retest data including resting-state and 7 different task-fMRI (see Methods).

The impact of the decomposition into connectivity modes and subsequent reconstruction of FC patterns is assessed in different scenarios. For all 7 fMRI tasks and resting-state, we find the existence of optimal reconstructions that maximize identifiability of functional connectomes. At those optimal solutions, edgewise identifiability as measured by intra-class-correlation is largely enhanced. The possible influence of motion (absolute frame displacement per session) in identifiability is assessed and found to be significant for all fMRI tasks but not for resting-state. We propose a generalization of this framework for cross-sectional data based on splitting the fMRI time-series into two halves and evaluate it on all four resting-state sessions. We assess the impact of different lengths of fMRI runs on identifiability. We also find that optimal PCA reconstruction leads to more robust associations to task-related behavioral measurements across visits (test-retest). Finally, we map task-specific functional edges as measured by intra-class correlation and identify which within and between functional networks (FNs) are the most task-specific.

We conclude by discussing the interpretation of the concept of brain connectivity modes, or eigenmodes in the functional connectivity domain. We make considerations on possible driving factors (mainly motion and task performance) that may limit the maximization of identifiability. Finally, we discuss the limitations of our study and future work and potential applications of this methodology.

## Materials and Methods

**Dataset.** The fMRI and behavioral dataset used in this work is from the Human Connectome Project (HCP, http://www.humanconnectome.org/), Release Q3.

**Data availability.** The dataset analyzed during the current study, i.e. the Human Connectome Project, Release Q3, is available in the Human Connectome Project repository (http://www.humanconnectome.org/). The processed functional connectomes obtained from this data and used for the current study are available from the corresponding author on reasonable request. Below follows the full description of the acquisition protocol and processing steps.

**HCP data.** We assessed the 100 unrelated subjects (54 females, 46 males, mean age = 29.1 ± 3.7 years) as provided at the HCP 900 subjects data release [16,17]. This subset of subjects provided by HCP ensures that they are not family relatives. This criterion was crucial to exclude the need of family-structure co-variables in our analyses as well as possible identifiability confounds. Per HCP protocol, all subjects gave written informed consent to the Human Connectome Project consortium. The fMRI resting-state runs (HCP filenames: rfMRI_REST1 and rfMRI_REST2) were acquired in separate sessions on two different days, with two different acquisitions (left to right or LR and right to left or RL) per day [18,19]. The seven fMRI tasks were the following: gambling (tfMRI_GAMBLING), relational (tfMRI_RELATIONAL), social (tfMRI_SOCIAL), working memory (tfMRI_WM), motor (tfMRI_MOTOR), language (tfMRI_LANGUAGE, including both a story-listening and arithmetic task) and emotion (tfMRI_EMOTION). The working memory, gambling and motor task were acquired on the first day, and the other tasks were acquired on the second day [17,20]. The HCP scanning protocol was approved by the local Institutional Review Board at Washington University in St. Louis. All experiments were performed in accordance with relevant guidelines and regulations. For all sessions, data from both the left-right (LR) and right-left (RL) phase-encoding runs were used to calculate connectivity matrices. Data acquisitions for each subject and for each task were tagged as test and retest (also referred as visits). In order to avoid confounds between test-retest and phase encoding, runs were evenly distributed on test-retest along the subjects. Hence, for half of the subjects LR was used as test and RL as retest and for the other half RL was used as test and LR as retest. This operation was done for all 7 fMRI tasks. For the case of resting-state, this procedure was done for both REST1 and REST2 separately. Full details on the HCP dataset have been published previously [17–19].

**Brain atlas.** We employed a cortical parcellation into 360 brain regions as recently proposed by Glasser et al. [21]. For completeness, 14 sub-cortical regions were added, as provided by the HCP release (filename "Atlas_ROI2.nii.gz"). To do so, this file was converted from NIFTI to CIFTI format by using the HCP workbench software [18,22] (http://www.humanconnectome.org/software/connectome-workbench.html, command *-cifti-create-label*)

**HCP preprocessing: functional data.** The HCP functional preprocessing pipeline [18,19] was used for the employed dataset. This pipeline included artifact removal, motion correction and registration to standard space. Full details on the pipeline can be found in [18,19]. The main steps were: spatial ("minimal") pre-processing, in both volumetric and grayordinate forms (i.e., where brain locations are stored as surface vertices [19]); weak highpass temporal filtering (> 2000s full width at half maximum) applied to both forms, achieving slow drift removal. MELODIC ICA [23] applied to volumetric data; artifact components identified using FIX [24]. Artifacts and motion-related time courses were regressed out (i.e. the 6 rigid-body parameter time-series, their backwards-looking temporal derivatives, plus all 12 resulting regressors squared) of both volumetric and grayordinate data [19].

For the resting-state fMRI data, we also added the following steps: global gray matter signal was regressed out of the voxel time courses [13]; a bandpass first-order Butterworth filter in forward and reverse directions [0.001 Hz, 0.08 Hz] [13] was applied (Matlab functions *butter* and *filtfilt*); the voxel time courses were z-scored and then averaged per brain region, excluding outlier time points outside of 3 standard deviation from the mean, using the workbench software [22] ( workbench command *-cifti-parcellate* ). For task fMRI data, we applied the same above mentioned steps except the bandpass filter, since it is still unclear the connection between different tasks and optimal frequency ranges [25].

Pearson correlation coefficients between pairs of nodal time courses were calculated (MATLAB command *corr*), resulting in a symmetric connectivity matrix for each fMRI session of each subject. Functional connectivity matrices were kept in its signed weighted form, hence neither thresholded nor binarized. Finally, the resulting individual functional connectivity matrices were ordered (rows and columns) according to 7 functional cortical sub-networks (FNs) as proposed by Yeo and colleagues [26]. For completeness, an 8th sub-network including the 14 HCP sub-cortical regions was added (as analogously done in recent paper [27]).

**PCA reconstruction of the individual connectivity profiles**

Principal component analysis (PCA) is a statistical procedure [28] that transforms a set of observations of possibly correlated variables into a set of values of linearly uncorrelated variables called principal components (or sometimes, principal modes of variation), ranked in descending order of explained variance of the initial data. PCA has been widely used for exploratory analysis of the underlying structure of data in many areas of science, from pattern recognition in genetics [29], to denoising and compression in image processing [30], and recently to dynamic functional connectivity patterns [31].

In this study we explored the use of PCA (MATLAB command *pca*) in the connectivity domain for improving the individual fingerprint in functional connectomes from a group-level perspective. The procedure starts by matching the number of principal components included with the number of functional connectomes of the dataset. By definition, this PCA-based decomposition accounts for 100% of the variance in the data. As mentioned above, components are ranked according to explained variance in descending order. The next step consists of the reconstruction of the individual functional connectomes as a function of the number of components included (see methodological scheme at Fig. 1). The rationale behind this analysis is that high-variance components might carry cohort-level functional connectivity information, whereas lower-variance components might carry subject-level functional connectivity information and, finally, lowest-variance components might carry noisy or artifactual connectivity information. By doing this iterative fine-grained exploration of number of components used, we are able to identify in a data-driven fashion what are the boundaries for group-level, individual-level, and artifactual-level components. That is, once extracted the main connectivity-based principal components (PCs), each individual connectivity profile is reconstructed based on its mean and the linear combination of the chosen PCs (see Fig. 1).

For 16 fMRI sessions (REST1 and 7 fMRI-tasks, with test-retest for each), we explored the property of individual fingerprinting for different levels of reconstruction based on the number of ranked components used. We kept the 2 additional fMRI runs from REST2 as validation set. In the next section we will define the function employed for evaluating the level of individual fingerprint at any reconstruction level.

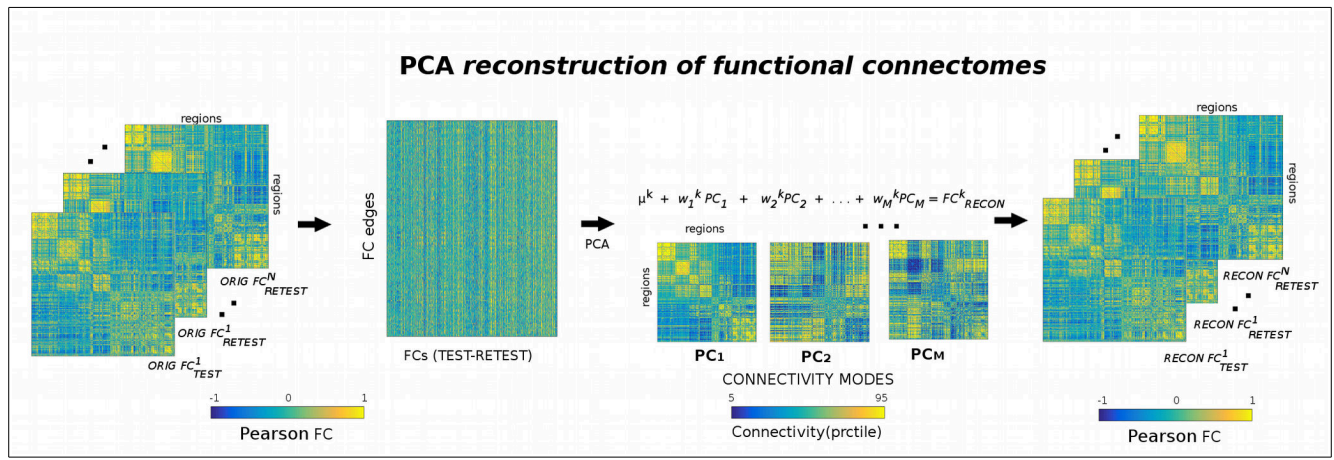

**Figure 1. Workflow scheme of the group-level principal component analysis (PCA) reconstruction procedure of individual functional connectomes (FC).** The upper triangular of each functional connectivity matrix (two FCs per subject, test-retest) is vectorized and added to a matrix where columns are sessions and rows are their vectorized functional connectivity patterns. Data are first centered: this is obtained by subtracting

the mean µ$^k$ from each column (where k goes from 1 to N subjects). Second, the PCA algorithm extracts the M principal components (i.e. the functional connectivity modes) associated to the whole population and their relative weights across subjects. The M orthogonal connectivity modes are then used to reconstruct back the FC of each subject (µ$^k$ is added back to the data). Colorbars indicate positive (yellow) to negative (blue) connectivity values: Pearson's correlation coefficient in the case of individual FC matrices (left and right sides of scheme), and unitless connectivity weights in the case of PCA FC-modes. For ease of visualization, the FC-modes colorbar ranges from 5$^{th}$ to 95$^{th}$ percentile of the distribution of values.

**Differential identifiability quality function**

The maximization of a functional connectome fingerprint relies on the assumption that the connectivity profiles should be, overall, more similar between visits or sessions of the same subject than between different subjects. Finn et al. [6] showed that, to a great extent, it is possible to robustly identify the functional connectome of a subject "target" from a sample database of FCs, simply by computing the spatial (Pearson) correlation of the target FC against the database ones. For identification, they used a set of connectivity matrices from one session for the database and connectivity matrices from a second session acquired on a different day as the target set. Then, given a query connectivity matrix from the target set, they computed the correlations between this matrix and all the connectivity matrices in the database. Finally, for each query, the predicted identity was picked as the one with the highest correlation coefficient, and assigned a score of 1 if the predicted identity matched the true identity, and a score of 0 otherwise. The success rate of this identification procedure was above 90% for resting-state sessions, and ranged between 54% and 87% when including task-task and task-rest sessions [6].

In order to evaluate our framework and the identifiability capability as a continuum, we generalized the above mentioned binary score system [6] to a more continuous score on the level of individual fingerprinting present on a set of test-retest functional connectomes. Hence we introduce the concept of "level of identifiability" on a set of functional connectomes.

Let **A** be the "identifiability matrix", i.e. the matrix of correlations (square, non symmetric) between the subjects' FCs test and retest. The dimension of **A** is $N^2$, where N is the number of subjects in the database. Let $I_{self}=<a_{ii}>$ represent the average of the main diagonal elements of **A**, which consist of the Pearson correlation values between visits of same subjects: from now on, we will refer to this quantity as self-identifiability or $I_{self}$. Similarly, let $I_{others}=<a_{ij}>$ define the average of the off-diagonal elements of matrix **A**, i.e. the correlation between visits of different subjects. Then we define the differential identifiability (**I$_{diff}$**) of the population as the difference between both terms:

$$I_{diff} = (I_{self} - I_{others}) * 100 \ , \ i \neq j \qquad (1)$$

which quantifies the difference between the average within-subject FCs similarity and the average between-subjects FCs similarity.

The higher the value of $I_{diff}$, the higher the individual fingerprint overall along the population. By defining $I_{diff}$, the optimization problem of differential identifiability is then reduced to maximizing $I_{diff}$. This consists of exploring within a range of number of components (M), and of finding the optimal number of components, m*, in the PCA decomposition that provides the maximum value of $I_{diff}$, namely $I_{diff}$*, for which:

$$I_{diff}* = \arg\max_{m \in M} I_{diff}(m) \qquad (2)$$

**Influence of the number of fMRI volumes on identifiability**

The effect of scan duration on individual measurements of functional connectivity has been well documented [15,32–34] and might be considered as a potential confound for the identifiability analysis presented here. We assessed, for REST1, the impact of the number of fMRI volumes in the PCA reconstruction and subsequent identifiability. To do so, we tested on shortening the session to different session-lengths corresponding to the following numbers of fMRI volumes: 5, 10, 25, 50, 100 and up to 1,200 in steps of 100. For each number of fMRI volumes, a full exploration of PCA reconstruction was run in order to obtain optimal number of components and subsequent identifiability score. This allowed us to assess the impact of PCA optimal reconstruction on different number of fMRI volumes.

**Edgewise subject identifiability and edgewise task identifiability**

The function defined earlier provides a way to quantify the differential identifiability at a whole-network level. We also quantified the edgewise subject identifiability by using intraclass correlation [35,36](ICC). ICC is a widely used measure in statistics, normally to assess the percent of agreement between units (or ratings/scores) of different groups (or raters/judges) [37]. It describes how strongly units in the same group resemble each other. The stronger the agreement, the higher its ICC value. We used ICC to quantify to which extent the connectivity value of an edge (functional connectivity value between two brain regions) could separate within and between subjects. In other words, the higher the ICC, the higher the identifiability of the connectivity edge.

Following the same rationale, one can also compute edgewise ICC when tasks are "raters" and "scores" are given by subjects. In this case, the higher the ICC, the more separable the different tasks across subjects and consequently the higher the task identifiability of the connectivity edge.

**Influence of motion regressors on self identifiability**

Finally, we tested if optimal self identifiability (i.e. same subject's FC similarity) after PCA reconstruction was linked to motion estimators. HCP data collection provides an estimate of average absolute (i.e. displacement from initial frame, file name "Movement_AbsoluteRMS_mean.txt") and frame-to-frame displacement (file name "Movement_RelativeRMS_mean.txt") for each run and fMRI session. We evaluated linear and $\log_{10}$-linear trends between self identifiability values after reconstruction and subject's motion displacement values (maximum between the two runs).

**Robustness in the association of functional connectivity with task-related scores before and after optimal PCA reconstruction.**

For each task fMRI session (test and retest, for both original and reconstructed functional connectomes), we computed the corresponding association matrices with the average response time and performance accuracy across the task (file pattern {TASK_name}_stats.txt). To do so, the edgewise (across subjects) Pearson's correlation between functional connectivity and the task feature was computed, producing a square correlation matrix of the same size of an FC matrix. Correlations in association matrices denote the correlation between functional connectivity values between two regions across subjects and a task feature (in our case, response time and accuracy)  Note that we excluded the MOTOR task (information was not available) and we only computed average response time in those tasks where accuracy was not clearly defined (i.e. SOCIAL and GAMBLING tasks). For each task-measurement, this produced four association matrices (based on original and reconstructed FC test, based on original and reconstructed FC retest). Each association matrix was filtered based on two steps. First, only significant correlations were kept (p<0.01) and second, only the functional associations forming the giant component were kept (as analogously done in Network Based Statistics or NBS, [38]).  We measure the robustness of the filtered association matrices by obtaining their intersection between test and retest, for the original and the reconstructed data. After this procedure, level of robust associations is measured by the remaining number of functional edges, and of nodes.  We also tested the significance of these numbers over null distributions created by randomly shuffling the task-measures 1,000 times, and computing the same two quantities (i.e. the overlapping number of nodes and number of edges of the giant component across the two visits). This procedure is similar to the random shuffling of group-membership done in NBS [38].

**Influence of PCA reconstruction on functional connectome-based predictive modeling**

Next, we evaluated the predictability of fluid intelligence (gF, mean = 16.2 ± 4.6) based on resting-state, as analogously done by Finn et al. [6]. We used connectome-based predictive modeling (CPM, [39]) to assess whether the predictability of fluid intelligence increases after optimal PCA reconstruction. CPM takes as input data connectivity matrices (FC here) and a vector of behavioral measures (gF here).  Validation of predictability was based on leave-one-out cross-validation [6]. Iteratively, edges of all subjects except one are associated to the behavioral measure using Pearson's correlation. Significant edges (p<0.01) with positive correlation with behavior are then summed into a single subject value (strength). A linear predictive model is then fitted to predict behavior (dependent variable) from strength values (independent variable). Finally, the strength of the subject left out is used on the linear model to obtain the corresponding predicted behavioral score [6,39].

We used REST1-AVG (i.e. average FCs from runs 1 and 2 of day 1) to test gF predictability before and after optimal PCA reconstruction. Edges with positive significant correlations with gF (p<0.01) were first selected. Significant edges were sorted based on their correlation with gF, weighted by their corresponding intraclass correlation (see Figure 3B1). Including ICC ensures the use of edges with a high individual fingerprint and hence allows for a better generalization of the predictive model. Finally, we built a predictive model in an additive fashion, by adding the most important 5, 10 up to the best 100 edges in steps of 5. At every step, gF predictability (i.e. the correlation between predicted and observed gF values) was computed before and after reconstruction.

# Results

The dataset used for this study consisted of functional data from the 100 unrelated subjects in the Q3 release of the HCP [16,17]. For each subject, we estimated 18 functional connectivity matrices: 4 corresponding to resting-state ( conventionally named REST1 test-retest and REST2 test-retest, see Methods), 14 corresponding to each of the 7 tasks (test-retest, where {TASK_NAME}_LR and {TASK_NAME}_RL are evenly distributed; see Methods). The multimodal parcellation used here, as proposed by Glasser et al. [21], includes 360 cortical brain regions. We added 14 subcortical regions, hence producing functional connectome matrices (square, symmetric) of 374 x 374 (see Methods for details).

For each task (including resting-state), individual functional connectomes (including two visits, test-retest, per subject) were reconstructed based on PCA by iteratively including different number of components. This procedure can be summarized as follows (Fig. 1): first, the upper triangular part of each individual functional connectivity matrix was vectorized and added to a matrix where columns are the subjects and rows are their full connectivity pattern; second, the PCA algorithm extracted the principal components (PCs) associated to the whole population; third, these components were projected back in the individual subjects' space, leading to a "reconstructed" version of each original connectivity profiles (Fig. 1).

From the test-retest pool of 100 unrelated subjects (total of 200 FC matrices per task and 200 per resting-state, for this experiment only the REST1 test-retest FCs were considered), a bootstrap technique was used to accurately estimate $I_{diff}$ for each value m of number of components, m={2,5,10:10:160}. This was meant to avoid results driven by a small subset of the population. To do so, 100 random samples comprising the test-retest pairs of 80 subjects (total of 160 FC matrices) were performed for each value of m.

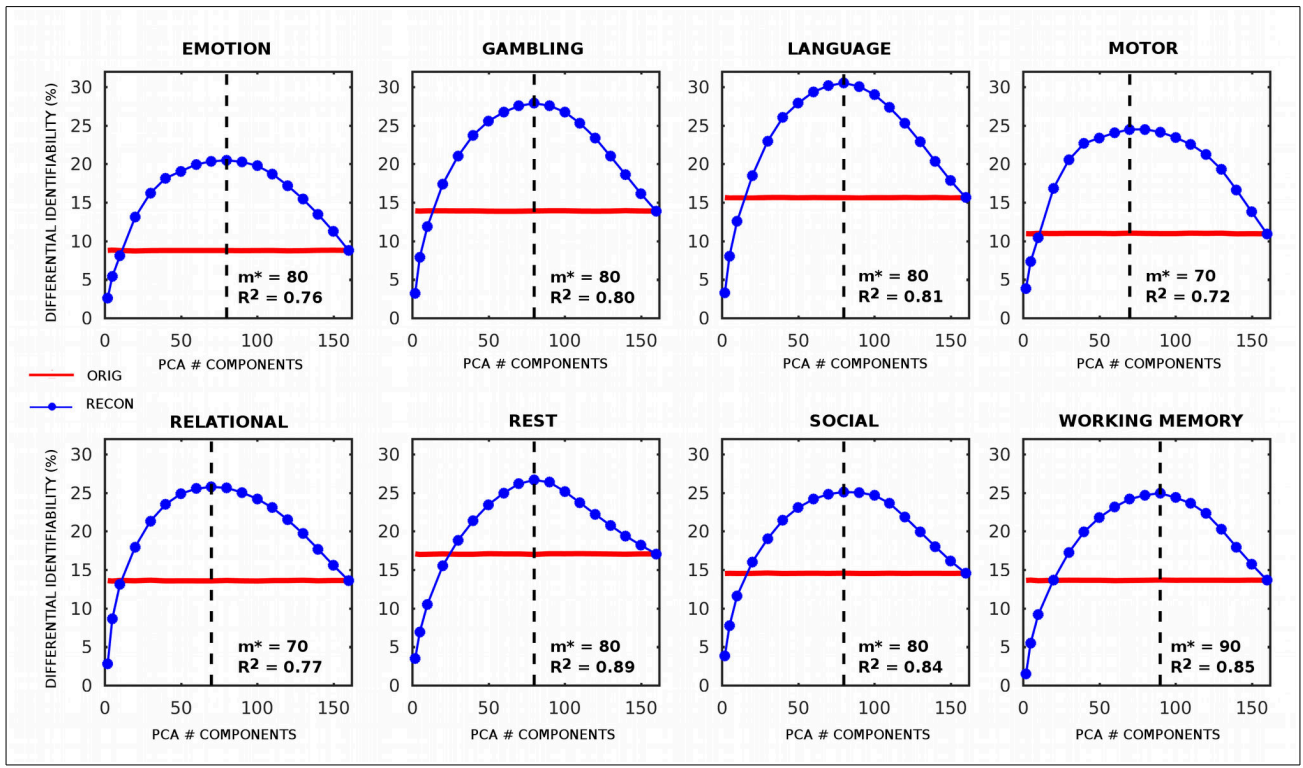

**Figure 2. Percent difference of the differential identifiability ($I_{diff}$) as a function of the number of PCA components used for reconstruction in resting-state and 7 fMRI tasks.** Plots show, for each task, the differential identifiability as a function of the number of PCA components used for reconstruction (evaluated at 2, 5, and 10 to 160 components in steps of 10). Red line denotes $I_{diff}$ for the original FCs, whereas blue line with circles denotes the identifiability for reconstructed FCs based on the different number of components sampled. For each subplot, the optimal number of components that maximizes $I_{diff}$ (m*) and the corresponding explained variance ($R^2$) are shown. The PCA reconstruction was tested on the resting state (REST) and 7 different task sessions provided by the HCP data (EMOTION, GAMBLING, LANGUAGE, MOTOR, RELATIONAL, SOCIAL, WORKING MEMORY, see Methods for details). To test the stability of the method, $I_{diff}$ was evaluated over 100 different runs. At each run, 80 subjects were randomly sampled from the HCP data pool of 100 unrelated subjects, 2 sessions for a total of 160 FCs at every run). The standard deviation of $I_{diff}$ (not shown in the plots) across runs was always lower than 0.8 %, for all the sessions considered, for original and reconstructed data.

## FC differential identifiability increases at optimal PCA reconstruction

For each number m of components retained for reconstruction, we tested the differential identifiability ($I_{diff}$) (see Methods) of the reconstructed connectomes with respect to the original ones, based on the following assumption: the connectivity profiles should be, overall, more similar between visits of the same subjects than between different subjects. For simplicity, in the main text we will show results for resting-state (REST) and for the MOTOR task . Results for all the other six tasks are shown in Fig. S1 and Fig. S2.  For all seven tasks and for resting-state, the PCA reconstructed functional connectomes outperformed the original ones in terms of $I_{diff}$ for a wide range of m (Fig. 2, Fig. S1 and Fig. S2). Each condition shows a slightly different optimal m* for maximizing $I_{diff}$ (e.g. m*=80 PCs for resting-state, m*=70 for the MOTOR task, Fig. 2). For all cases, the variance of the functional data kept in the reconstruction was between 72% and 89%. We also compared original and optimally reconstructed FC data by using identification rate (as proposed by [6]). As shown in Fig. S4, we found that identification rate was higher after optimal reconstruction for all fMRI tasks and for resting-state. When assessing identifiability rates on REST1-AVG (two-runs of day 1) and REST2-AVG (two-runs of day 2), even higher rates were obtained at the optimal PCA reconstruction. In particular, when REST1-AVG was the target, identifiability rate increased from 94% to 98%, whereas when REST2-AVG was the target it increased from 93% to 97%.

By definition, $I_{diff}$ optimization is constrained to the availability of test-retest fMRI sessions, which is not common in many acquired fMRI experiments, especially in cross-sectional assessments of clinical populations. To cover the necessity of assessing individual fingerprinting in these scenarios, we computed "two-halves" individual FCs by splitting each of the 4 resting state sessions in two parts (~600 fMRI volumes each). We then evaluated $I_{diff}$ before and after reconstruction, with "test-retest" sessions now being the first and the second part of the same fMRI acquisition. Again, the PCA reconstructed functional connectomes keep outperforming the original ones for a wide range of m (see Fig. S3). We next explored further properties of the optimal PCA reconstruction for REST and the MOTOR task.

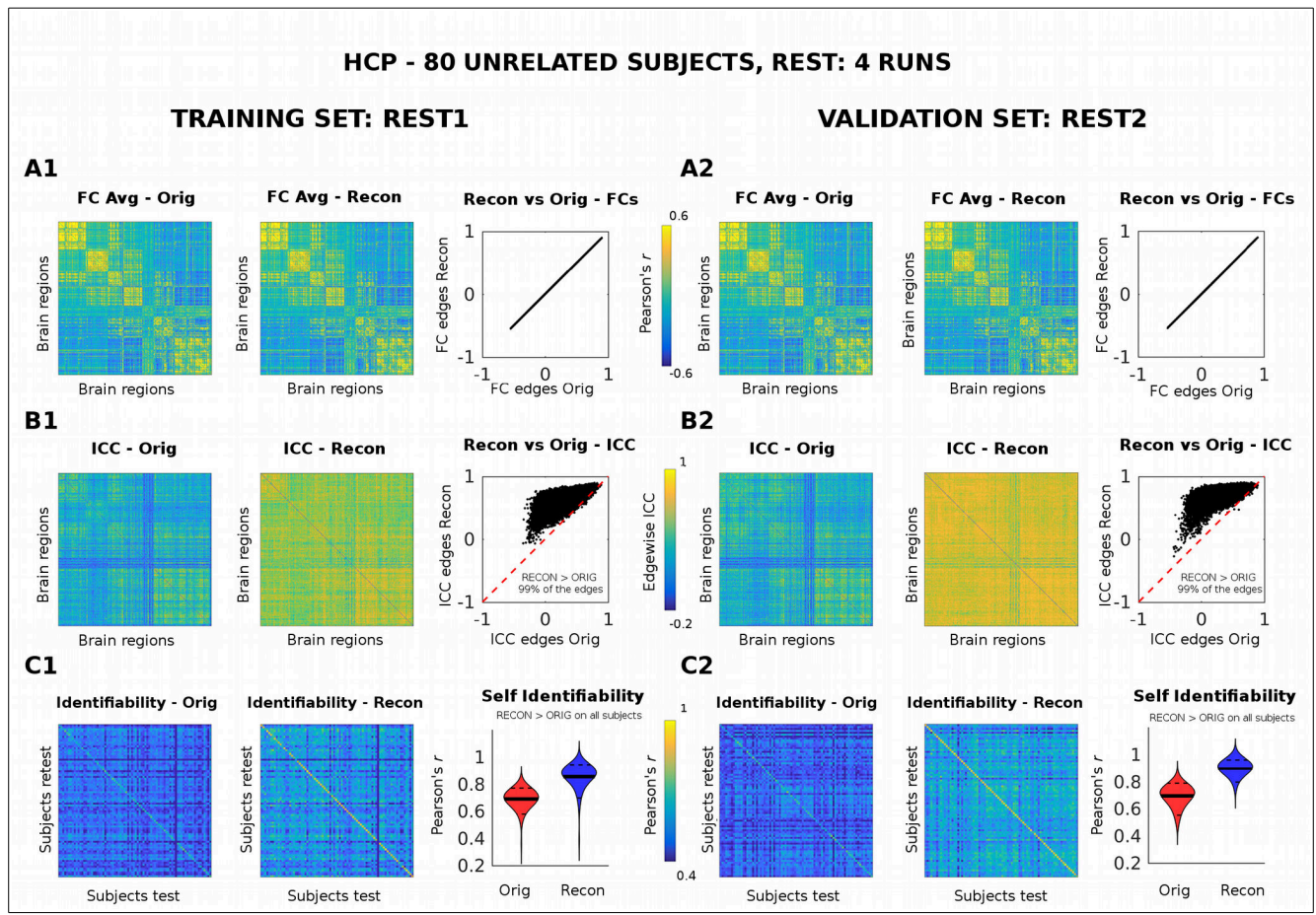

**Figure 3. Evaluation and validation of PCA reconstruction on resting-state functional connectomes (FCs) at the optimal point (m\* = 80).** The resting FCs of 80 subjects were reconstructed by using group-level PCA. The optimal number of PCA components was 80 (see Fig. 2). Results shown correspond to a single run. The PCA reconstruction was first evaluated on the REST1 (test and retest, training set, left panel). The FC-modes extracted from the training set were then used to reconstruct two different resting state sessions, namely

REST2 test and retest (here used as a test set, right panel). **A1-A2)** From left to right: the group averaged FC of the original data; the group averaged FC of the reconstructed data; the scatter plot edge by edge of the reconstructed group averaged FC (y axis) vs original group averaged FC (x axis). **B1-B2)** From left to right: the intra-class correlation (ICC), computed over each FC edge, for the original data; the edgewise ICC for the reconstructed data; the scatter plot edge by edge of the reconstructed ICC values (y axis) vs original ICC values (x axis). The inset reports the percentage of edges where ICC increased after reconstruction (black dots on top of the red dashed line) from those that did not (black dots below of the red line). **C1-C2)** From left to right: Identifiability matrix (i.e. Pearson's correlation coefficient between functional profiles across subjects and sessions, see Methods) of the original data; identifiability matrix of the reconstructed data; violin plot of the "self identifiability" (i.e. the main diagonal of the identifiability matrix) distribution across the 80 subjects, for original (Orig, red) and reconstructed (Recon, blue). The solid black lines depict the mean value of the distribution; the dashed black lines the 5 and 95 percentiles. Note that, as specified by the inset, the self identifiability of each subject always improves after PCA reconstruction, both for the training and test set.

## Reconstructed Functional connectomes: resting state

For the REST data, we tested the goodness of this method at the optimal PCA reconstruction point with respect to the original FCs in three different ways: 1) by comparing the group average functional connectomes before and after optimal PCA reconstruction; 2) by computing the edgewise intraclass correlation (ICC, i.e. how good a single edge can separate different subjects, see Methods), before and after optimal PCA reconstruction; 3) by evaluating $I_{diff}$ (as described in Methods), before and after optimal PCA reconstruction.

The 4 resting-state fMRI acquisitions per HCP subject available allowed us to perform evaluation and validation of our methodology as follows: the optimal PCA reconstruction was first evaluated on the "training set" consisting of REST 1 (Fig. 3, left panel), at the optimal number of 80 PCA components (as depicted in Fig. 2). The 80 FC-modes extracted from the training set were then used as the "orthogonal connectivity basis" through which reconstruct the two other FCs of the same subjects, i.e REST2 (Fig. 3, right panel).

Both for training and validation sets, the optimal PCA reconstruction preserves the main characteristics of the functional connectomes (the group averaged functional connectomes before and after reconstruction are almost identical, Fig. 3 A1, A2). Nonetheless, the edgewise ICC largely increase after optimal reconstruction for almost all edges (99%), both in the training and test cases (Fig. 3 B1, B2). In accordance with results shown in Fig.2, the self identifiability of the subjects' FCs after reconstruction increases on all subjects (Fig. 3 C1, C2).

## Reconstructed Functional connectomes: motor task

Evaluation of the proposed methodology was also performed for all seven tasks available in the HCP dataset. We show as example results for the MOTOR task at the optimal reconstruction (m*= 70, Fig. 4). Also, Fig. S2 summarizes results for all the other tasks. Even in this case the reconstructed group average FC resembles the original one almost perfectly (Fig. 3A); edgewise ICC improves after optimal reconstruction on 96% of the functional edges (Fig. 3B); self identifiability increases after reconstruction on all subjects (Fig. 3C), with some subjects showing a larger increase than others (as shown by the violin plot distributions of self identifiability after reconstruction, Fig. 3C).

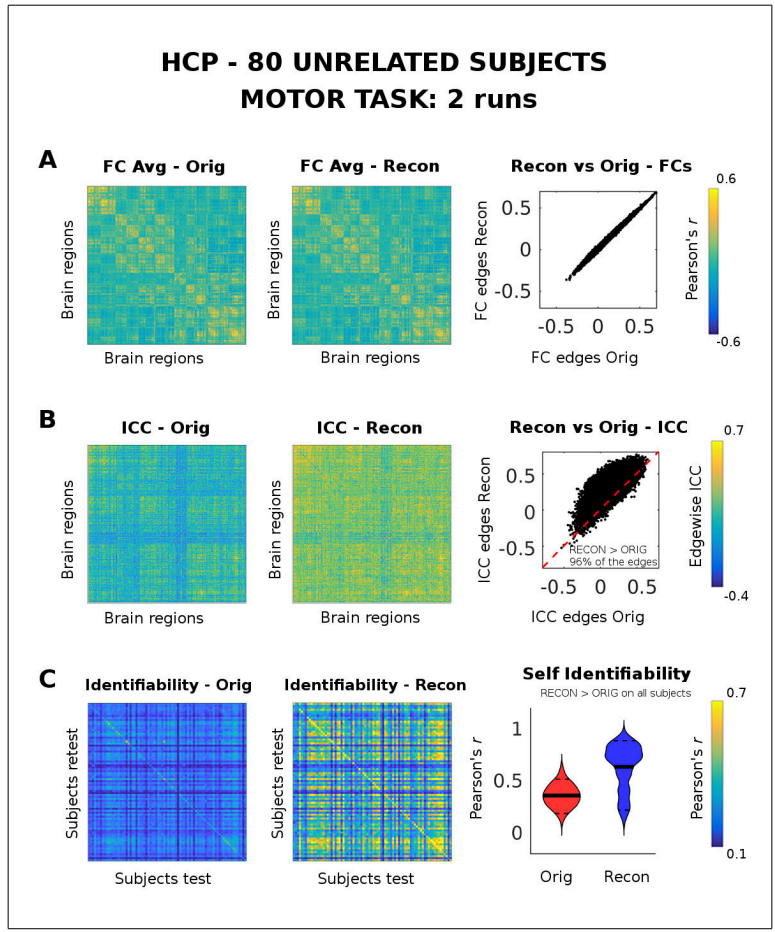

**Figure 4. Evaluation of optimal PCA reconstruction (m* = 70) on motor task-based functional connectomes (FCs) .** The motor task-based FCs of 80 subjects were reconstructed at the optimal number of PCA components (i.e. 70, see Fig.2), for 1 single run. The PCA reconstruction was evaluated on the using the two runs available per subject. **A1)** From left to right: the group averaged FC of the original data; the group averaged FC of the reconstructed data; the scatter plot edge by edge of the reconstructed group averaged FC (y axis) vs original group averaged FC (x axis). **B1)** From left to right: the intra-class correlation (ICC), computed over each FC edge, for the original data; the edgewise ICC for the reconstructed data; the scatter plot edge by edge of the reconstructed ICC values (y axis) vs original ICC values(x axis). The inset reports the percentage of edges where ICC increased after reconstruction (black dots on top of the red dashed line) from those that did not (black dots below of the red line). **C1)** From left to right: Identifiability matrix (i.e. the Pearson's correlation between functional profiles between subjects and sessions, see Methods) of the original data; identifiability matrix of the reconstructed data; violin plot of the "self identifiability"(i.e. the main diagonal of the identifiability matrix) distribution across the 80 subjects, for original (Orig, red) and reconstructed (Recon, blue). The solid black lines depict the mean value of the distribution; the dashed black lines the 5 and 95 percentiles. Note that self identifiability of each subject always improves after PCA reconstruction.

## Effect of the number of time frames on differential identifiability

Next, we evaluated the impact of the length of the fMRI session on the reconstruction of functional connectomes. We estimated the trend of the differential identifiability on the REST1 session, when including different number of fMRI volumes in the FC estimation, and compared the curve obtained after PCA reconstruction to the one obtained from the original data (see Fig. 5). Results from this analysis can be interpreted in two ways: the PCA optimal reconstruction always improves identifiability with respect to the original FC data. Remarkably, the level of identifiability obtained originally by using the whole length of REST1 is outperformed in the reconstructed data when only using 200 fMRI volumes (Fig. 5).

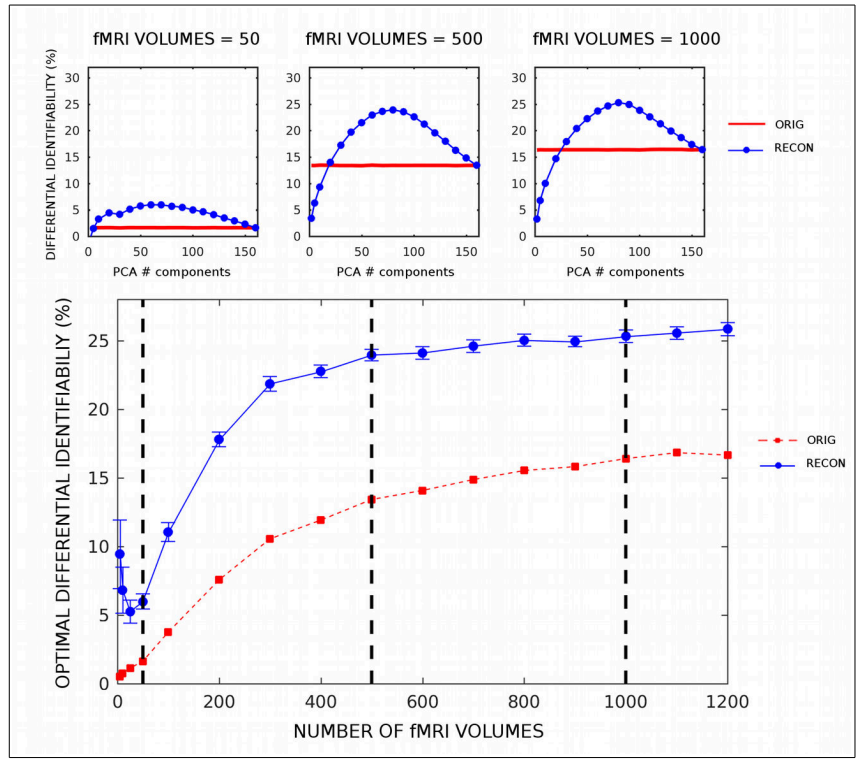

**Figure 5. Optimal differential identifiability ($I_{diff}$) as a function of the number of fMRI volumes used for reconstruction on REST1. Bottom:** plot shows the optimal average $I_{diff}$ across 100 runs (similarly to Fig. 2 and Fig. S3, see Methods for details), as a function of the number of fMRI volumes (frames) used for FC evaluation and subsequent reconstruction. Values tested are: 5, 10, 25, 50 frames, and from 100 to 1,200 frames in steps of 100. Red line with circles denotes the average $I_{diff}$ for the original FCs (standard deviation across runs were always < 0.4 % and are not shown in the plot), whereas blue line with circles denotes the identifiability for reconstructed FCs based on the different number of fMRI volumes retained (blue vertical bars indicate standard deviation across runs). **Top:** insets show the $I_{diff}$ curves (blue line, reconstructed $I_{diff}$; red line, original $I_{diff}$; same as Fig.2 and S3) per optimal number of PCA components (evaluated at 2, 5, and 10 to 160 components in steps of 10) for three different choices of number of fMRI frames retained (50, 500 and 1,000 fMRI volumes respectively). The three vertical dashed black lines in bottom plot show the correspondent optimal $I_{diff}$ value for each of three insets on top.

## PCA reconstruction improves robustness and reproducibility of associations with behavior

An important question regarding this approach is to what extent it increases the association of the reconstructed connectomes with behavior and/or task performance. Ideally, improving the identifiability of functional connectomes should increase the robustness and reproducibility of the associations with behaviour. That is, if a functional connectivity subsystem is associated with task performance in one experiment (test), one would expect it to be connected to it in a repetition of the experiment (retest).

| | | FC ORIGINAL | | FC PCA RECONSTRUCTED | |
| --- | --- | --- | --- | --- | --- |
| | | test ∩ retest | test ∩ retest | test ∩ retest | test ∩ retest |
| **Task Name** | **Perf. Score** | # edges (prctile) | # nodes (prctile) | # edges (prctile) | # nodes (prctile) |
| Emotion | Accuracy | 4 (66%) | 4 (68%) | 0 (0%) | 0 (0%) |
| Emotion | Response Time | 24 (96%) | 17 (95%) | 96 (98%) | 47 (98%) |
| Gambling | Response Time | 6 (32%) | 5 (32%) | 8 (26%) | 6 (24%) |
| Language | Accuracy | 2 (32%) | 2 (32%) | 26 (98%) | 21 (98%) |
| Language | Response Time | 10 (71%) | 8 (66%) | 26 (83%) | 20 (82%) |
| Relational | Accuracy | 24 (92%) | 20 (94%) | 110 (98%) | 75 (99%) |
| Relational | Response Time | 114 (99%) | 48 (99%) | 304 (99%) | 99 (99%) |
| Social | Response Time | 0 (%) | 0 (%) | 6 (47%) | 6 (51%) |
| W. Memory | Accuracy | 28 (94%) | 26 (96%) | 80 (98%) | 58 (99%) |
| W. Memory | Response Time | 18 (85%) | 13 (81%) | 36 (88%) | 31 (91%) |

**Table 1. Robustness in the association of functional connectivity with task-related scores before and after optimal reconstruction.** For each task, score and visit (test and retest), an association matrix correlating the scores with all pairwise functional vectors is computed. Values are thresholded by statistical significance (p<0.01) and the giant component of the thresholded association matrix is obtained. Finally, the intersection of the giant components of test and retest is obtained and considered a robust association. Each row of the table indicates the resulting robust association in terms of number of functional edges, number of nodes involved. This procedure is repeated with the original FC data and with the optimally PCA reconstructed data. Percentile values between parenthesis account for respective null models where the performance score is randomly shuffled (each distribution is built upon 1,000 repetitions). Note how, with the exception of EMOTION with accuracy, the reconstructed connectomes outperform the original ones across all tasks and scores.

Briefly, for each task-measurement, association matrices were obtained, filtered by significance and by giant component. We measured the robustness of the filtered association matrices by obtaining their intersection between both visits for the original and for the reconstructed data. Robustness is measured by the remaining number of functional edges, and of nodes. We also tested the significance of these numbers over null distributions (see Methods for details). Results per task-measurement for original and reconstructed FC data are reported in Table 1. Apart from one exception (EMOTION, accuracy), the optimally reconstructed connectomes outperformed the original ones for robust associations across all tasks in having more functional edges, more nodes (or brain regions) and in being in higher percentiles with respect to their null models. Overall, these results indicate that PCA optimal reconstruction displays more robust associations to task-related measurements across visits.

**PCA reconstruction improves connectome-based predictive modeling of behavior**

We assessed the impact of this approach on predictive modeling from FCs to behavior. Based on connectome-based predictive modeling (CPM, [39]), we tested fluid intelligence (gF) predictability on REST1-AVG (see Methods for details), as analogously done by Finn et al. [6].

Fig. 6 shows predictability of gF for up to the top 100 most relevant edges (added iteratively), sorted by correlation with gF (weighted by ICC) , for original and reconstructed connectomes. Note how PCA reconstructed connectomes outperform the original in gF predictability at every step of the curve (Fig. 6).

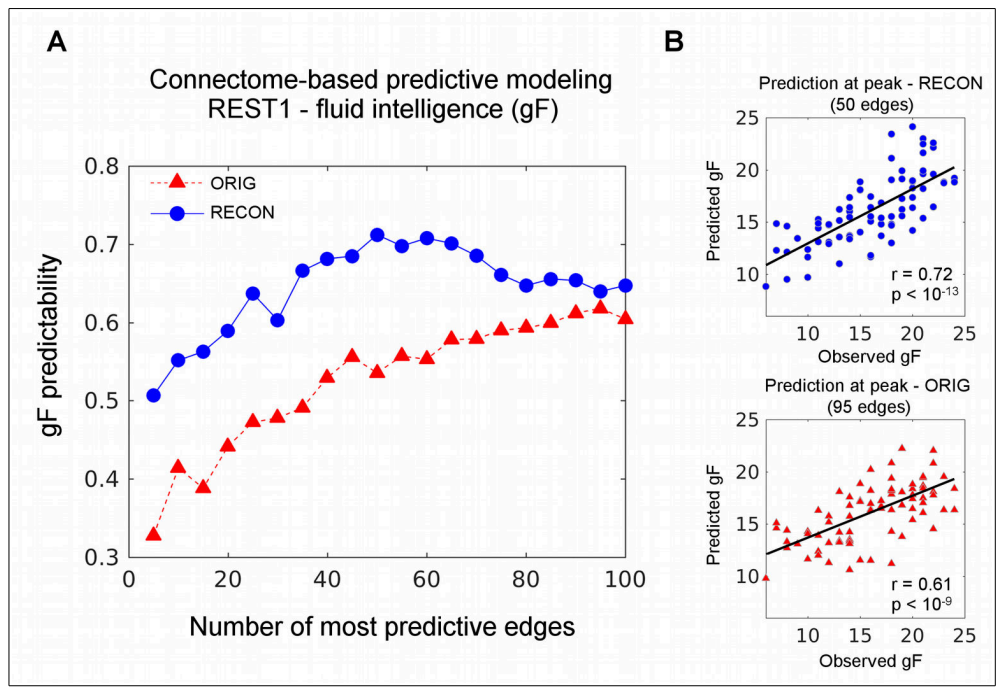

**Figure 6: Predictive modeling of fluid intelligence before and after optimal PCA reconstruction. A)** predictability level of fluid intelligence (gF) from resting state FCs when selected up to the top 100 most predictive edges. Blue curve represents the prediction obtained from reconstructed FCs (RECON), whereas red curve is from original data (ORIG). **B)** Scatter plots of optimal predictions found. Top: observed versus predicted gF values for RECON (blue circles, r=0.72); bottom: observed versus predicted gF values for ORIG (bottom, red triangles, r=0.61).

## Self identifiability in tasks correlates with motion displacement

Further inspection of self identifiability of subjects reveals differences, mainly between resting-state and the fMRI tasks. The violin plot distributions of self identifiability values after reconstruction for resting are unimodal, whereas task sessions display bimodal and sometimes even trimodal shapes (Fig. 3C, Fig. S2). This suggests that subject's identifiability might relate to the goodness of his test-retest FCs. That is, if one or more of the individual sessions is highly compromised or corrupted (due to motion or other factors), it might become very challenging to improve the similarity between the FCs of the same subject. Indeed, self identifiability values after reconstruction for the task sessions negatively correlate with the mean absolute motion displacement from the first fMRI frame (AbsoluteRMS, see Methods), particularly with the maximum AbsoluteRMS between the two sessions (Fig. S5). That is, the higher the presence of motion in at least one the sessions of the subject, the lower the subject identifiability. Notably, the log-linear trend between identifiability and AbsoluteRMS is evident across all the different tasks, but it is not present for the resting-state session (Fig. S5). Moreover, no significant correlation between self identifiability and task response time nor between identifiability and task accuracy (see Methods) was observed for any task (results not shown).

**Reconstructed Functional connectomes: edgewise subject/task identifiability**

Finally, we tested the edgewise identifiability properties of the optimally reconstructed functional connectomes, on all subjects and tasks, in two different ways. One, by computing the edgewise subject identifiability over all tasks (i.e., intraclass correlation across subjects, see Methods). In other words, identifying which specific set of functional connections can separate between subjects regardless the tasks (resting state included). The other, by computing the edgewise task identifiability over all subjects (i.e. intraclass correlation across tasks, see Methods); that is, which specific functional connections can separate between tasks (resting state included) regardless which subjects are performing them (see Methods). Interestingly, there are sub-networks that are associated to subject and task identifiability more than others, both in terms of within- and between- FNs (Fig. 7). Particularly, edges involving connectivity within the visual and dorsal attentional networks, as well as edges between those two networks, appear among the top 5 sub-networks highly implicated in task identifiability (Fig. 7 D,E,F). The fronto-parietal and DMN networks and its between-network connectivity with the attentional networks is highly involved in subject identifiability (Fig. 7 A,B,C). Both Subject ICC and Task ICC results were consistent and qualitatively similar before and after optimal PCA reconstruction (results not shown for original FC data).

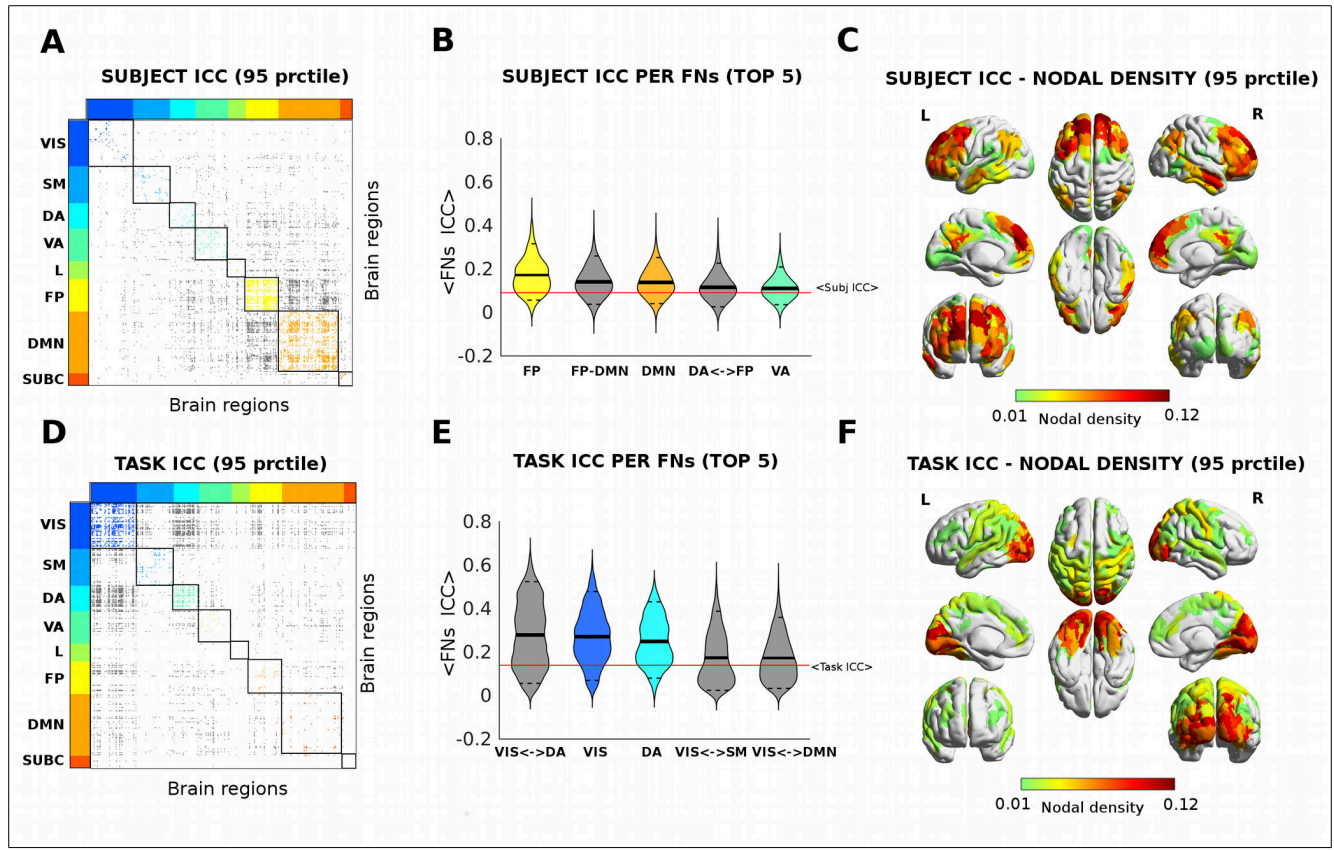

**Figure 7. Intra-class correlation (ICC) analysis of subject identifiability and task identifiability. A-D) Edgewise ICC for identifiability.** Figure shows functional connections for which ICC values were significantly higher (> 95$^{th}$ percentile of the distribution). The brain regions are ordered according to Yeo's (Yeo et al., 2011) functional resting state networks (FNs): Visual (VIS), Somato-Motor (SM), Dorsal Attention (DA), Ventral Attention (VA), Limbic system (L), Fronto-Parietal (FP), Default Mode Network (DMN), and subcortical regions (SUBC). The colored dots depict ICC value across subjects in different within FNs networks; gray dots indicate significant ICC edges between FNs. The most prominent networks for subject's identifiability (6A) appear to be: FP, DMN and the interaction FP-DMN and with the attentional networks. For task identifiability (6B): VIS, and the interaction VIS-DA; DA, SM and the VA. **B-E) Violin plot of edgewise ICC for the top 5 FNs.** The edgewise ICC distribution per within or between FNs interaction, for the five with the highest ICC value for identifiability. Each different color indicates a different within FN (as in 6A-D), while the gray indicates ICC values between functional networks. The solid black lines depict the mean value of the distribution; the dashed black lines the 5 and 95 percentiles; the solid red line indicates the whole-brain mean ICC value. **C-F) Brain render of ICC subject identifiability as nodal density per region.** The strength per brain region computed as sum of edges above the 95 percentile threshold divided by the total number of edges per region gives an assessment of the overall prominence of each brain region for subject's and task identifiability. Note how, the occipital lobe is prominent in both task and subject identifiability, while frontal areas show higher nodal density for subject's ICC, as opposed to dorsal areas in task ICC.

# Discussion

The neuroscientific community is advancing towards the era of large public data repositories (such as the Human Connectome Project [16], or the 1,000 Functional Connectomes Project [40], the era of reproducibility of brain data and neuroscientific results [41], and the exciting avenue of linking large-scale brain connectivity profiles to single subject's genetic [42], clinical, demographical and behavioral features [5,39]. In this respect, improving the reliability and robustness of individual fingerprinting in the connectivity domain (both functional and structural) is a crucial next step in the area of brain connectomics.

We here presented a framework that addresses this point from a cohort-level perspective. We used principal component analysis to decompose and optimally reconstruct functional connectomes obtained from the 100 unrelated subjects cohort (two sessions, test-retest) from the HCP benchmark, both in resting state and all seven fMRI-tasks. Results indicate that this method improves, on a data-driven fashion, both the global and the local (edgewise) individual fingerprint (as measured by identifiability) of the functional connectivity profiles, independently from the acquired task.

PCA is a method commonly used to provide a simpler representation of the data at hand, by compressing most of the variance of the data in a reduced number of orthogonal components, or eigenvectors. For instance, in face recognition problems, the retained eigenvectors ("eigenfaces") are used to denoise the initial images and improve the identification of the face in the image [43]. Similarly, we here mapped and ranked the principal connectivity modes from a set of functional connectomes. Hence, by maximizing differential identifiability, the reconstructed functional connectomes provided a denoised or more accurate version of the original ones.

The simplicity of the approach allowed us to test the optimal number of eigenmodes or components to retain for an optimal differential identifiability of the functional connectomes (Fig. 2) across different tasks. Also, when splitting the resting-state time-series into two halves, each resting state run shows a similar optimal m* (m*=80 PCs for split data, Fig. S3;

m*=80 for full data, Fig. 2). Also, resting-state is the session which shows greater improvement after optimal PCA reconstruction (Fig. 3) in terms of edgewise ICC and self identifiability. Moreover, the improvement in the individual identification was also substantial for all the task sessions (Fig. 4 and Fig. S2). Despite subtle differences between $I_{diff}$ and identification rate as proposed by [6], results suggest that maximizing $I_{diff}$ leads to an increased identification rate for all tasks and resting-state sessions (Fig. S4).

Furthermore, in order to evaluate how differences in the number of fMRI volumes acquired relates to identifiability, we tested the optimal identifiability for REST1 before and after reconstruction when changing the number of frames available in the evaluation of functional connectomes (Fig. 5). This result is noteworthy for two reasons. First, it shows that differential identifiability after reconstruction always improves, also with low number of frames (Fig. 5). Second, and more importantly, the identifiability rate obtained from the original data using the whole resting-state acquisition (1,200 volumes) is achieved after optimal PCA reconstruction when only considering 200 frames of the same acquisition. Indeed, given the asymptotic behavior of the identifiability on the original data, values near 18% or above seem almost unreachable for any acquisition length without the optimal reconstruction. This striking result suggests that the proposed method may have an impact in helping reducing and optimizing the duration and design of fMRI protocols, procedures and costs. Further research could refine the assessment of trends and relations between acquisition length and identifiability.

Another crucial issue worth to tackle here was related to the association of the reconstructed patterns with behavioral scores of the subjects. Indeed, a method with the aim of improving identifiability needs to be tested on the association with cognitive-behavioral scores. Ideally, if two functional connectomes of a same subject have better identifiability after reconstruction across fMRI runs, one would expect to observe or detect more robust and stable associations with behavior across the two visits. Motivated by this hypothesis, we tested whether the functional connectomes obtained after PCA decomposition increases the robustness and reproducibility of their association with behavioral scores, here identified as average response time and average task accuracy (see Table 1 and Methods). With a single exception, the results confirm the hypothesis that this method contributes to obtaining stronger, more reliable and more robust associations with behavioral scores across runs. We also tested whether this method could improve the prediction of individual behaviour from functional connectomes. As a test-bed, we used fluid intelligence to show that the resting state reconstructed connectomes outperform the original ones in the prediction of fluid intelligence (Fig. 6), showing evidence of the potential of this approach for associations between brain connectivity and behavior. Altogether, these findings suggest that this method provides a step forward towards reproducibility and robustness in brain connectomics as well as for linking connectivity profiles to cognitive "ground-truth" or behavioral scores.

We found head motion (as measured by absolute frame displacement, see Methods) to be significantly associated with differential identifiability (negatively correlated) for all tasks and, interestingly, in a minor way for resting-state (Fig. S5). This might lead to two main considerations. One, that in resting state the principal components discarded may be successfully carrying most of the motion-induced artifacts left over in the functional connectome domain. The second, that motion during resting-state sessions might be more homogeneous between subjects or better isolated in the discarded components (ranked by explained variance, see discussion on limitations below) than when the subject is cognitively engaged in a task that include time-controlled events and interactions. Ultimately, the more brain connectivity modes retained, the more fine-grained information (i.e. variance explained) is kept in the functional connectomes after optimal reconstruction. However, this comes at the risk of carrying over session-specific and/or motion-induced connectivity artifacts, not beneficial for identifiability.

We extended the question to whether it is possible to identify a subject or a task performed based solely on characteristic functional connectivity patterns. Our results based on intraclass correlation show some functional networks (FNs) specialization, being some FNs more involved in edgewise task-identifiability and others in subject-identifiability (Fig. 7). This is in line with recent studies reporting that individual differences in many tasks can be stable trait markers [44,45], as well as that individual fingerprint is not homogeneous across FNs [6,45]. We extend these questions on individual fingerprinting by showing here that identifiability across individuals and across tasks might overlap. Indeed, we found prominent FNs-based connections sensitive to both subjects and tasks being performed (Fig. 7).

This study has some limitations. The optimal number of components is dataset dependent (e.g., size of the cohort, heterogeneity within the cohort, acquisition and processing characteristics) and cannot be easily extrapolated from one dataset to another one. Also, here we are selecting the principal brain modes based on the ranking by variance explained (as it is the normal procedure in PCA). Different selection of the optimal subset of PCs should be explored (i.e by using simulated annealing [46]).

This work adds up to the emerging new field of features extraction in brain connectomics (independent subsystem detection through independent component analysis [27,47]; dimensionality reduction and connectome denoising through PCA here), that can contribute to the association of neuroimaging data with clinical/genetic biomarkers, as well as to the exploration of the underlying latent structures and factors present in the connectome architecture of the human brain. Future studies should explore more advanced models of features extraction as well as the connections of these denoised connectivity profiles with behavior/performance/cognition. For instance, on cross-sectional studies one could try to find optimal number of components that allows for mapping demographics or cognitive performance variables. Another exciting avenue of further investigation relates to the predictability of new (healthy or clinical) functional connectivity profiles, once given the connectivity modes obtained from a "benchmark" cohort. Future work should also explore, with more advanced approaches such as machine learning techniques, up to what extent on the reconstructed connectomes it is possible to identify a subject with higher accuracy than using original FC. Other interesting avenues also involve the application of this methodology to structural connectivity patterns, and the dependence of the reconstruction on aging through longitudinal analyses.

Individual fingerprinting within the functional connectivity domain is a critical attribute for further research on brain connectomics. Here, we used data from the Human Connectome Project to demonstrate that the optimal reconstruction of the individual FCs through connectivity eigenmodes maximizes subject identifiability across resting-state and all seven tasks evaluated. The subject identifiability of the reconstructed individual connectivity profiles increased both at the global and edgewise level, also when the reconstruction was imposed on additional sessions of the subjects. We extended this approach to also map the most task-sensitive functional connections. Results showed that is possible to maximize individual fingerprinting in the functional connectivity domain regardless of the task, a crucial next step in the area of brain connectivity towards individualized connectomics.

## Acknowledgments

Data were provided [in part] by the Human Connectome Project, WU-Minn Consortium (Principal Investigators: David Van Essen and Kamil Ugurbil; 1U54MH091657) funded by the 16 NIH Institutes and Centers that support the NIH Blueprint for Neuroscience Research; and by the McDonnell Center for Systems Neuroscience at Washington University. This work was partially supported by NIH R01EB022574,by NIH R01MH108467, and by the Indiana Clinical and Translational Sciences Institute (UL1TR001108) from the NIH, National Center for Advancing Translational Sciences, Clinical and Translational Sciences Award. We would like to thank Dr. Olaf Sporns, Dr. Alex Fornito, Dr. Gesualdo Scutari and Dr. Mario Dzemidzic for insightful discussions.

## Author contributions

E.A and J.G conceptualized the study, processed the MRI data, designed the framework and performed the analyses, interpreted the results and wrote the manuscript.

## Declaration of interests

The authors declare no competing interests.

# The quest for identifiability in human functional connectomes

Enrico Amico[1,2] and Joaquín Goñi [1,2,3,*]
[1] School of Industrial Engineering, Purdue University, West-Lafayette, IN, USA
[2] Purdue Institute for Integrative Neuroscience, Purdue University, West-Lafayette, IN, USA
[3] Weldon School of Biomedical Engineering, Purdue University, West-Lafayette, IN, USA

# Supplementary Information

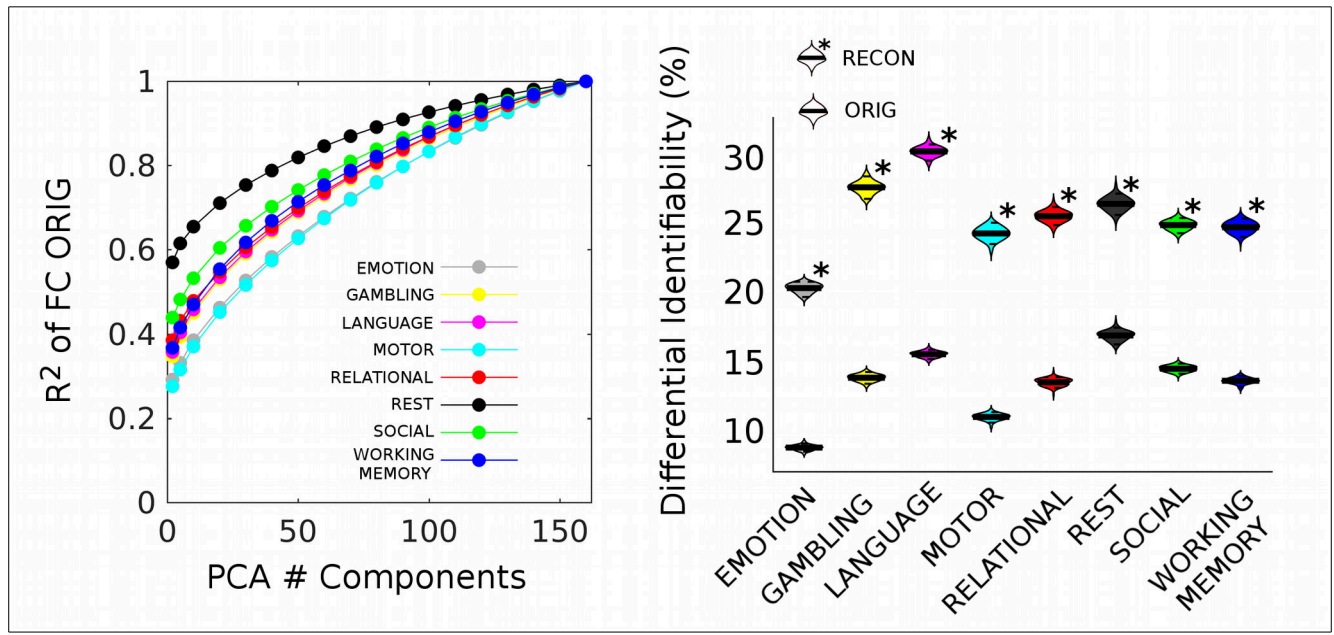

**Figure S1. Explained variance and differential identifiability ($I_{diff}$) across sessions.** Left: the variance explained (R-square) of the original data from the PCA reconstruction, for different number of PCA components employed. Each session is plotted with a different color. Right: violin plots show the distribution of the FC individual identifiability (see Methods) across subjects, for each fMRI session (each one has a distinct color), before and after PCA reconstruction. The solid black lines of the violins depict the mean value of the distribution. The asterisk indicates the individual identifiability distributions after reconstruction. Note how the PCA reconstruction always improves the individual identifiability.

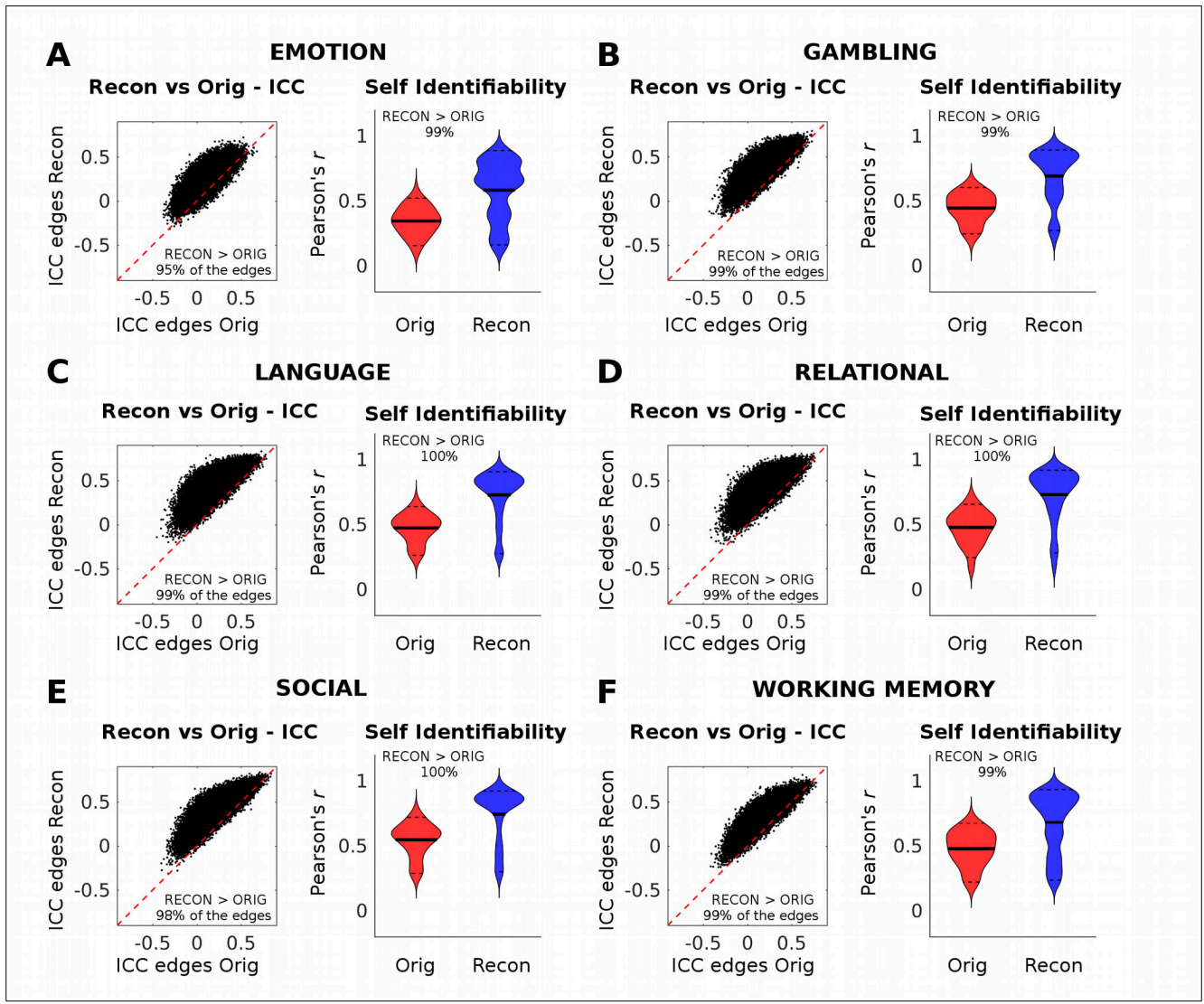

**Figure S2. Summary of results on ICC and self-identifiability ($I_{self}$) for the six fMRI tasks not shown in the main text.** Left: the scatter plot edge by edge of the reconstructed ICC values (y axis) vs original ICC values (x axis). The inset reports the percentage of edges where ICC increased after reconstruction (top of the red dashed line) from those that did not (low of the red line). Right: violin plot of the "self identifiability" (i.e., the main diagonal of the identifiability matrix, see Methods) distribution across the 80 subjects, for original (ORIG, red) and reconstructed (RECON, blue). The solid black lines depict the mean value of the distribution; the dashed black lines the 5 and 95 percentiles. The inset specifies the percentage of subjects whose identifiability has improved after PCA reconstruction.

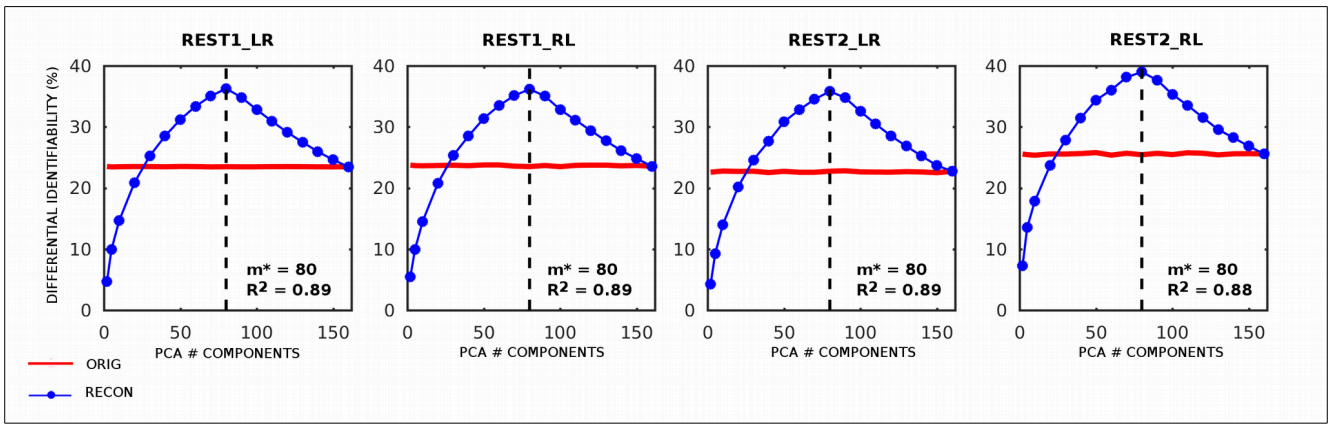

**Figure S3. Percent difference of the differential identifiability ($I_{diff}$) as a function of the number of PCA components used for reconstruction in split resting-state sessions.** Plots show, for each split resting state sessions (test = first 600 fMRI frames, retest = second 600 fMRI frames, see Methods for details), $I_{diff}$ as a function of the number of PCA components used for reconstruction (evaluated at 2, 5, and 10 to 160 components in steps of 10). Red line denotes $I_{diff}$ for the original FCs, whereas blue line with circles denotes the identifiability for reconstructed FCs based on the different number of components sampled. For each subplot, the optimal number of components that maximizes differential identifiability (m*) and the corresponding explained variance (R2) are shown. To test the stability of the method, $I_{diff}$ was evaluated over 100 different runs. At each run, 80 subjects were randomly sampled from the HCP resting-state data pool of 100 unrelated subjects, 4 sessions (REST1_LR, REST1_RL, REST2_LR and REST2_RL) for a total of 160 FCs at every run. The standard deviation of $I_{diff}$ (not shown in the plots) across runs was always lower then 0.9 %, for all the sessions considered, for both original and reconstructed data.

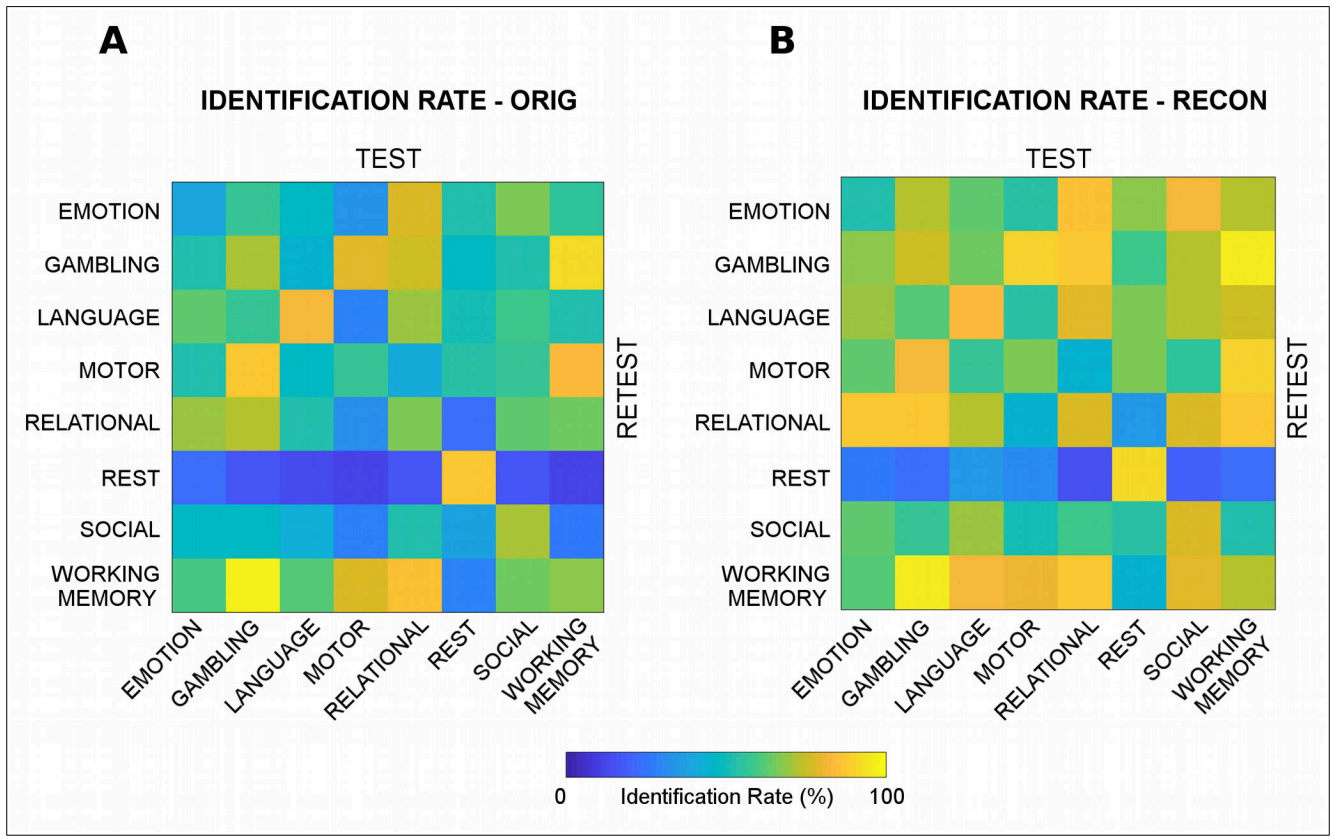

**Figure S4: Identifiability rates for all possible combinations of test-retest** (within the same sequence and between different sequences), before and after PCA reconstruction. Note that identifiability rates in the main diagonal (i.e. test-retest from the same sequence) were the average of test-retest and retest-test rates. Also, for main diagonal values, the optimal number of PCA components m* was based on the findings of Fig. 2. For off diagonal values, where FCs come from different tasks ($T_i, T_j$), optimal reconstruction on the mixed data matrix was defined as $m^*_{mixed} = \max(m^*_{Ti}, m^*_{Tj})$.

|  | EMOTION | GAMBLING | LANGUAGE | MOTOR | RELATIONAL | REST | SOCIAL | WORKING MEMORY |
|---|---|---|---|---|---|---|---|---|
| **Abs_RMS - μ ± σ** | 0.36 ± 0.39 | 0.39 ± 0.44 | 0.44 ± 0.45 | 0.49 ± 0.33 | 0.43 ± 0.46 | 0.82 ± 0.44 | 0.37 ± 0.32 | 0.56 ± 0.52 |
| **RT (ms) - μ ± σ** | 798 ± 143 | 418 ± 116 | 359 ± 350 | n.a. | 1762 ± 327 | n.a | 1103 ± 358 | 884 ± 149 |
| **ACC (%) μ ± σ** | 98 ± 4 | n.a. | 88 ± 9 | n.a. | 76 ± 14 | n.a. | n.a | 86 ± 10 |

**Table S1:** The summary statistic (mean and standard deviation across the 100 unrelated subjects) for the motion and behavioral variables employed in Fig.S5 and Table1, respectively, for each of the fMRI task and resting-state. In order from top to bottom row: absolute frame displacement (Abs_RMS, unitless); task response time (RT, milliseconds); task accuracy (ACC, percentage).

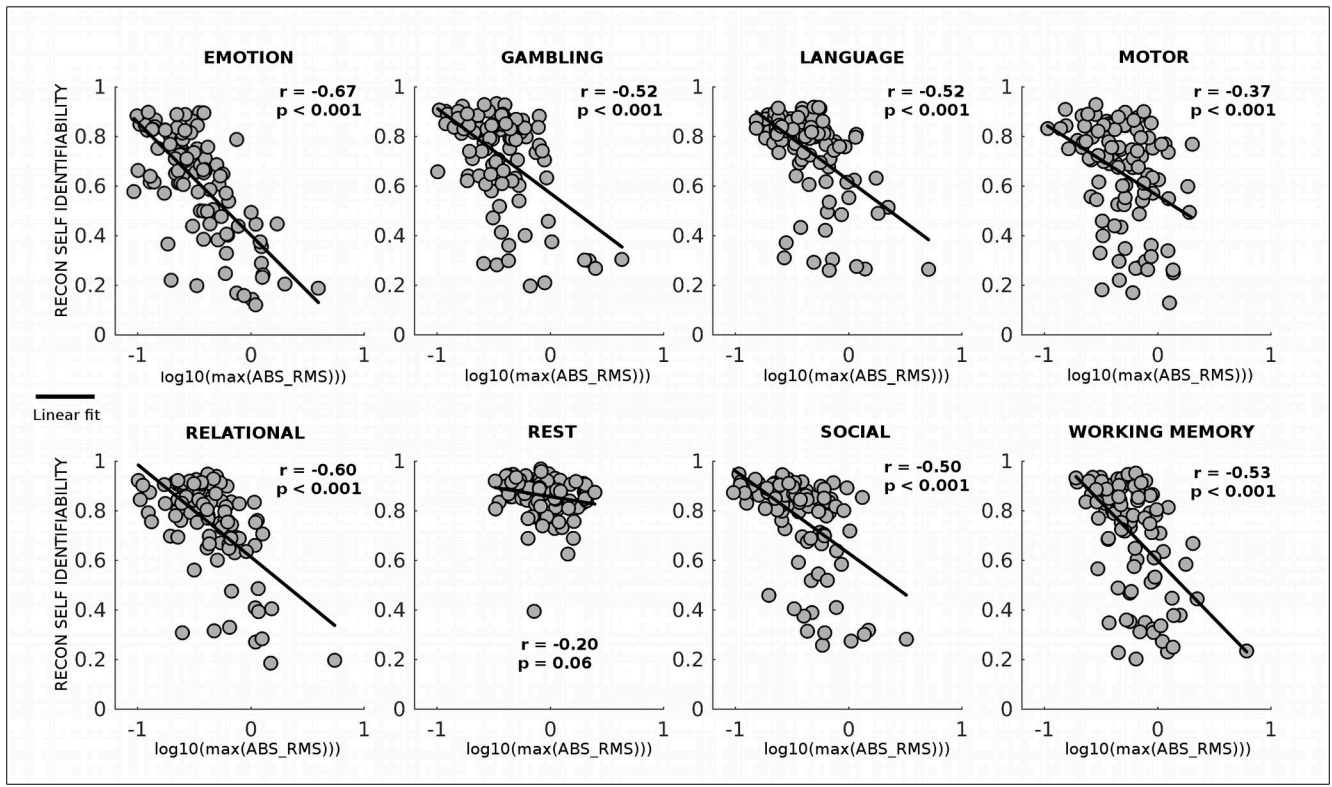

**Figure S5. Log-linear trend evaluation between self identifiability ($I_{self}$) and mean absolute frame displacement.** Plot shows, for each resting-state and task session, the scatter plot between individual self identifiability (see Methods for details) values after reconstruction (y-axis) and the $\log_{10}$ of the maximum value ( across the two sessions) of the average absolute frame displacement (ABS_RMS, x-axis). Solid lines show the linear fit of the scatter plots, and the insets report Pearson's correlation coefficient (r) between these two variables, with the associated significance (p-value). Note how there is a significant negative correlation (p<0.001) between increases in self identifiability and ABS_RMS across all tasks. No significant linear trend is present for the REST acquisition.